\documentclass[12pt,a4paper]{article}
\usepackage{graphicx}
\usepackage{amssymb,amsfonts,amsmath}
\usepackage{color}
\usepackage{cite}
\usepackage{bm}
\begin{document}

\newcommand{\EQ}{Eq.~}
\newcommand{\EQS}{Eqs.~}
\newcommand{\FIG}{Fig.~}
\newcommand{\FIGS}{Figs.~}
\newcommand{\TAB}{Tab.~}
\newcommand{\TABS}{Tabs.~}
\newcommand{\SEC}{Sec.~}
\newcommand{\SECS}{Secs.~}

\title{Immunization of networks with community structure}
\author{Naoki Masuda${}^{1,2}$
\\
\ \\
\ \\
${}^{1}$ 
Graduate School of Information Science and Technology,\\
The University of Tokyo,\\
7-3-1 Hongo, Bunkyo, Tokyo 113-8656, Japan
\ \\
${}^2$
PRESTO, Japan Science and Technology Agency,\\
4-1-8 Honcho, Kawaguchi, Saitama 332-0012, Japan\\
\ \\
masuda@mist.i.u-tokyo.ac.jp}

\setlength{\baselineskip}{0.77cm}
\maketitle

\newpage

\begin{abstract}
\setlength{\baselineskip}{0.77cm} 
In this study, an efficient method to immunize modular networks (\textit{i.e.}, networks with community structure) is proposed. The immunization of networks aims at fragmenting networks into small parts with a small number of removed nodes. Its applications include prevention of epidemic spreading, intentional attacks on networks, and conservation of ecosystems. Although preferential immunization of hubs is efficient, good immunization strategies for modular networks have not been established. On the basis of an immunization strategy based on the eigenvector centrality, we develop an analytical framework for immunizing modular networks. To this end, we quantify the contribution of each node to the connectivity in a coarse-grained network among modules. We verify the effectiveness of the proposed method by applying it to model and real networks with modular structure.
\end{abstract}

\newpage

\section{Introduction}\label{sec:introduction}

The spread of epidemics can be considered to
occur on networks that describe contacts
between individuals \cite{Newman03siam,Boccaletti06,Barrat08book}. The
size and dynamics of epidemics heavily depend on the structure of
the contact network. In particular, 
in networks in which the number of contacts
per individual (\textit{i.e.}, the degree)
is heterogeneous, as represented by scale-free
networks, epidemic spreading can occur on a large scale
even at a small infection rate
\cite{Hethcote84,Anderson86,May88trans,Cohen00prl,Callaway00prl,Pastor01prl}.

Given a limited dose of immunization, it is practically necessary to
establish efficient immunization strategies against epidemics
occurring on networks. If an appropriate ordering of immunization of
nodes in a network is followed, the potential risk of large-scale
epidemic spreading can be suppressed.  We measure the efficiency of
immunization by assessing the capability of the immunization strategy
to fragment the network into small parts with a small number of
sequentially removed nodes. This method of assessment has a wide
applicability beyond prevention of epidemics.  In ecology, it is
important to identify the nodes in a food web whose removal causes the
catastrophic disintegration of the food web, which would seriously
damage the ecosystem \cite{Sole01,Dunne02el}. The possibility of
efficient immunization of a network also implies that the network is
vulnerable to intentional attacks such as those attributable to terrorism
\cite{Albert00nat}.

The standard solution to an immunization problem is to immunize hubs
(\textit{i.e.}, nodes having large degrees) preferentially
\cite{Callaway00prl,Albert00nat,Cohen01prl}.
The degree-based immunization strategy and its variants
are very efficient for
scale-free network models and some real data
\cite{Callaway00prl,Albert00nat,Cohen01prl,Holme02pre_attack,Cohen03prl,Holme04epl}.
However, many real networks are more structured than 
merely having heterogeneous degree distributions.
Partly because of this factor, 
an immunization strategy based on a graph partition algorithm
performs better than degree-based strategies \cite{Chen08prl}.
Immunization strategies involving the preferential removal of
nodes with large betweenness centrality
(betweenness-based strategies; 
see \SEC\ref{sec:results} for definition) also perform better
than degree-based strategies in some networks 
\cite{Holme02pre_attack,Ueno08}. Developing
efficient immunization strategies for general complex networks
is an unresolved question.

In this study, we focus on networks with modular
structure.
By definition, nodes in a network with modular structure
are partitioned into multiple modules (also called communities) such that
the number of links connecting the nodes in the same module is
relatively large.
The number of links connecting different modules is relatively small.
Such networks abound in various fields
 \cite{Wasserman94,Girvan02,Newman04epjb,Fortunato09}.
In simple cases in which modules are homogeneous and of equal size,
epidemic dynamics \cite{Becker95,Ball02mb,Schinazi02tpb} and
immunization \cite{Becker95,Ball04mb,Ball06smmr}
have been mathematically analyzed
in the limit of the infinite network size.
However, in practical applications, relevant networks are finite,
modules in a network are heterogeneous in various aspects
\cite{Fortunato09}, and nodes in a module play different roles
\cite{Guimera05pnas,Guimera05nat,Guimera05jsm}.  Metapopulation
modeling is a promising approach to the understanding of epidemic
dynamics in such modular networks \cite{Colizza07plosm,Colizza08jtb}.
Establishing practical immunization strategies for general modular
networks is an important issue.

We develop an immunization strategy for 
modular networks by extending an analytical framework
proposed recently \cite{Restrepo08prl}. 
It is our contention that it is important to consider
the role of each node in the coarse-grained network among modules
rather than in the original network so as to
preferentially immunize nodes that bridge
important modules.
Some algorithms for community detection effectively solve the
same problem \cite{Girvan02,Fortunato09}.
We believe that our method is much less
computationally expensive than these methods and therefore is
suitable for large modular networks.

\section{Methods}

\subsection{Immunization based on dynamical importance of nodes}\label{sub:Restrepo}

Consider an undirected and unweighted contact network with $N$
nodes. Even though our results can be easily extended to the case of
weighted networks, this study is confined to the immunization
of unweighted networks 
for simplicity.  An immunization strategy is an ordering of all
the nodes in a network
according to which the nodes are removed.  The fraction of
the removed nodes is set equal to $1-p$ ($0\le p\le 1$); the fraction
of the remaining nodes is equal to $p$.  The fraction of nodes contained
in the largest connected component (LCC) is
denoted by $S$. In a good immunization strategy, $S$ 
is small with a
small number of removed nodes, \textit{i.e.}, with a large $p$.

Restrepo and colleagues proposed an immunization strategy based on the
so-called dynamical importance of nodes \cite{Restrepo08prl}.
Although the dynamical importance is defined for directed networks
\cite{Restrepo06prl}, the exposition of their results in this section
is concerned with the undirected version.
The adjacency matrix is denoted by $A$; $A_{ij}=1$ when
node $i$ and node $j$ are adjacent, and $A_{ij}=0$ otherwise.  Because
$A$ is a symmetric matrix, all the eigenvalues of $A$ are real.  The
largest eigenvalue of $A$ and the corresponding eigenvector, which is
called the Perron vector, are denoted by $\lambda$ and $\bm u$,
respectively. In networks with low clustering (\textit{i.e.}, small
density of triangles), the LCC is large (\textit{i.e.}, $S={\rm O}(1)$) if
and only if $\lambda$ exceeds unity \cite{Restrepo08prl}.  The Perron
vector $\bm u$ is the mode that survives after multiplying $A$
repeatedly to an almost arbitrary initial $N$-dimensional vector.
Intuitively, the multiplication of $A$ implies the spread of epidemics to
the nearest neighbors.  A large $\lambda$ implies the efficient expansion
of the LCC.

When $p<1$, we generate the effective adjacency matrix
from the network composed of only
the remaining nodes and the links among these nodes.
We apply the threshold condition to the effective adjacency matrix
to determine whether the LCC is large. In this way, we can estimate
the critical value of $p$ with regard to the percolation transition.


The dynamical importance of node $k$, denoted by $I_k$, is
defined by the decrement of $\lambda$ owing to the removal of node $k$.
The linearized eigenequation after removing node $k$ is expressed as
\begin{equation}
(A+\Delta A)(\bm u+\Delta \bm u)=(\lambda+\Delta\lambda)
(\bm u+\Delta \bm u),
\label{eq:perturb}
\end{equation}
where $(\Delta A)_{ij}=-A_{ij}(\delta_{ik}+\delta_{jk})$ and
$\delta$ is Kronecker's delta.
Because the $k$th element of $\bm u$, denoted by $u_k$,
necessarily becomes zero owing to the removal of node
$k$,
the appropriate perturbation is given by $\Delta \bm u =
\Delta^{\prime}
\bm u -
u_k\hat{e}_k$, where $\hat{e}_k$ is the unit vector for the $k$th
component and $\Delta^{\prime} \bm u$ is an $N$-dimensional 
small vector. By inserting these expressions and 
$A\bm u=\lambda\bm u$ into \EQ\eqref{eq:perturb},
we obtain the following equation to first order:
\begin{equation}
I_k \equiv -\frac{\Delta\lambda}{\lambda}
\approx \frac{u_k^2}{\sum_i u_i^2}.
\end{equation}
Therefore, $I_k$ for undirected networks is equal to the square of the
eigenvector centrality \cite{Bonacich72}.

In the immunization strategy developed by Restrepo et al.
\cite{Restrepo08prl}, which we label as the Res strategy,
we first 
remove the node with the largest $I_k$. Then, we recalculate the
dynamical importance of each node in the updated network
to determine the second node to be removed.
We repeat this procedure. This method works efficiently
in various networks \cite{Restrepo08prl}.

\subsection{Localized epidemics in modular networks}\label{sub:localized}

The threshold condition $\lambda>1$ is ineffective for modular
networks.  To demonstrate this, consider an ad hoc modular network
composed of $N_{\rm M}$ homogeneous modules of equal size $N/N_{\rm M}$.  A node
is connected to each of the $N/N_{\rm M}-1$ nodes in the same module with
probability 1 and to each of the $N-(N/N_{\rm M})$ nodes in the other
modules with probability $\epsilon$. A small value of $\epsilon$
implies modular structure of the network \cite{Girvan02,Fortunato09}.
We can approximate the adjacency matrix $A$ by the following 
block-circulant matrix composed of $N_{\rm M}\times N_{\rm M}$ blocks, each of which is
an $(N/N_{\rm M})\times (N/N_{\rm M})$ matrix. Let $E$ be the $(N/N_{\rm M})\times
(N/N_{\rm M})$ unit matrix, and $J$ be the $(N/N_{\rm M})\times (N/N_{\rm M})$ matrix
whose all elements are unity.  The $N_{\rm M}$ diagonal blocks of $A$ are
equal to $J-E$.  If we approximate the probability
that a link exists
between two nodes in different modules by the weight of the
link, which is not crucial for the following arguments, the
$N_{\rm M}(N_{\rm M}-1)$ off-diagonal blocks of $A$ are equal to $\epsilon J$.  An
example network in the case of
$N=8$ and $N_{\rm M}=2$ is shown in \FIG\ref{fig:community
anneal}.

The $N_{\rm M}$ leading eigenvalues of the approximated adjacency
matrix are represented by
\begin{equation}
\lambda_i = -1+\frac{N}{N_{\rm M}}
\left[1+
\frac{\epsilon\rho^i\left(1-\rho^{(N_{\rm M}-1)i}\right)}
{1-\rho^i}\right],\quad (1\le i\le N_{\rm M}),
\label{eq:lambda_i anneal}
\end{equation}
where $\rho$ is an $N_{\rm M}$th generic root of unity.
Although \EQ\eqref{eq:lambda_i anneal}
more simply indicates the existence of an
$(N_{\rm M}-1)$-fold degenerate eigenvalue $-1+(1-\epsilon)N/N_{\rm
M}$ and a nondegenerate eigenvalue $-1+N/N_{\rm M}+\epsilon(N_{\rm
M}-1)/N_{\rm M}$, we use \EQ\eqref{eq:lambda_i anneal}
for theoretical developments below.
For further analysis,
we fix a specific $\rho$.
The corresponding eigenvectors are given by
\begin{equation}
\bm u_i=
(\underbrace{1\;\ldots \;1}_{N/N_{\rm M} \mbox{ times}}\;
\rho^i\;\ldots\; \rho^i\; \rho^{2i}\;\ldots\; \rho^{2i}\;\ldots\;
\underbrace{\rho^{(N_{\rm M}-1)i}\;\ldots\; 
\rho^{(N_{\rm M}-1)i}}_{N/N_{\rm M} \mbox{ times}})^{\top},\quad
(1\le i\le N_{\rm M}),
\label{eq:u_i anneal}
\end{equation}
where $\top$ denotes the transpose.
The other $N-N_{\rm M}$ eigenmodes have degenerated 
eigenvalues $-1$ and are
irrelevant to the percolation transition.

When $\epsilon$ is small, $\lambda_1$, $\ldots$, $\lambda_{N_{\rm M}}$
are almost the same. In the limit $\epsilon\to 0$,
we obtain $\lambda_1=\ldots=\lambda_{N_{\rm M}}=N/N_{\rm M}-1$.
In this limit, 
a proper linear summation of $\bm u_i$ ($1\le i\le N_{\rm M}$)
yields a localized mode represented by
\begin{equation}
\bm u_i^{\prime}=
(0\;\ldots \;0\; 1\; \ldots \;1\;
0\;\ldots \;0)^{\top},\quad
(1\le i\le N_{\rm M}),
\label{eq:u_i anneal localized}
\end{equation}
where a block of ones appears from the $(((i-1)N/N_{\rm M})+1)$th element to
the $(iN/N_{\rm M})$th element.
Each $\bm u_i^{\prime}$ represents a mode that is localized 
in a module.

According to the criterion explained in \SEC\ref{sub:Restrepo}, the
LCC is large when any of the values $\lambda_1$, $\ldots$, $\lambda_{N_{\rm M}}$
exceeds unity. When there are more than two nodes in each module
(\textit{i.e.}, $N/N_{\rm M}>2$), the LCC
is large even in the limit $\epsilon\to 0$, because
$\lambda=N/N_{\rm M}-1$. However, when $\epsilon\to 0$,
the LCC does not extend beyond a single
module, \textit{i.e.},
$S\le N/N_{\rm M}$.  When there are many
modules (\textit{i.e.},
large $N_{\rm M}$), the result for $\epsilon\to 0$ implies that the actual $S$
is small. When $\epsilon>0$ is small,
a similar relation holds true. In this case,
the largest eigenvalue is not degenerated.
However, for a moderate $p$, 
the LCC tends to contain a majority of nodes
in a single module and does not extend beyond the module.
Such a LCC is regarded to be large by the Res strategy,
whereas it is actually small when $N_{\rm M}$ is large.

In summary, the Res strategy applied to modular networks may be
inefficient, because it does not distinguish between local and global
epidemics.  The same is the case for degree-based immunization
strategies in which
hubs are preferentially immunized. If a considerable number of
hubs contribute to intramodular but not to intermodular connectivity,
alternative
strategies may work better.
Even though we have dealt with networks with modules of 
equal size, the discussions above can also be applied to modular networks
in which the size of modules is heterogeneous.

\subsection{Module-based immunization strategy}

We develop an
immunization strategy that can be applied to modular networks.
By definition, intermodular links are rare
compared to intramodular links in a modular network.
If intermodular links are
preferentially
removed during immunization, the modular structure will be preserved
throughout the
immunization procedure.  Therefore, if the LCC at a certain value of
$p$ contains a considerable number of modules that are
connected at this value of $p$,
many nodes in each of such modules are likely to
belong to the LCC.  On this basis, for simplicity, we assume that all
the nodes in each module belong to the LCC or none of them belongs to
the LCC.
To establish an efficient immunization strategy for modular networks,
we apply the Res strategy
to the coarse-grained network representing the connectivity
among modules.

Given a partition of nodes into $N_{\rm M}$ modules,
we define an $N_{\rm M}\times N_{\rm M}$ coarse-grained adjacency matrix
$\tilde{A}$ as
\begin{equation}
\tilde{A}_{IJ}=(1-\delta_{IJ})
\sum_{i\in M_I, j\in M_J} A_{ij},\quad
(1\le I, J\le N_{\rm M}),
\end{equation}
where $M_I$ denotes the $I$th module. The matrix $\tilde{A}$ is weighted,
and $\tilde{A}_{IJ}$ is equal to the number of links between $M_I$ and
$M_J$. 
It should be noted that
$\tilde{A}_{II}$ is set to 0 to respect the assumption that all the nodes
in a module are simultaneously included in or excluded from the LCC.
Otherwise, a localized mode such as
$(1\; 0\; \ldots\; 0)^{\top}$ may become the Perron vector of
$\tilde{A}$, owing to which
epidemics restricted to a single module cannot be ruled out.

The Perron vector $\tilde{\bm u}=(\tilde{u}_1\; \ldots\;
\tilde{u}_{N_{\rm M}})^{\top}$ of $\tilde{A}$ is determined by
$\tilde{A}\tilde{\bm u}=\tilde{\lambda}\tilde{\bm u}$,
where $\tilde{\lambda}$ is the largest eigenvalue of $\tilde{A}$.
$\tilde{u}_i$ represents the importance
of the $i$th module in terms of the eigenvector centrality.

We calculate the shift in $\tilde{\lambda}$,
denoted by $\Delta\tilde{\lambda}$, owing to the removal of a single
node $k$.  We denote the index of the module that node $k$ belongs to
by $K$. The removal of node $k$ elicits a change in the 
coarse-grained adjacency matrix by 
\begin{equation}
(\Delta\tilde{A})_{IJ}=
-\delta_{IK}d_{kJ}-\delta_{JK}d_{kI} + 2\delta_{IK}\delta_{JK}d_{kK},
\label{eq:Delta A coarse}
\end{equation}
where $d_{kI}$ is the number of
intermodular 
links that exist between node $k$ and module $M_I$, \textit{i.e.},
\begin{equation}
d_{kI}\equiv \sum_{i\in M_I} A_{ki}.
\end{equation}
It should be noted that $\Delta\tilde{A}_{KK}=0$.

To calculate $\Delta\tilde{\lambda}$,
it is necessary to evaluate the amount of perturbation 
in $\tilde{\bm u}$ owing to the node removal.
Generally, $\tilde{u}_K$ is perturbed by an amount larger than
$\tilde{u}_I$ ($I\neq K)$ because only the elements of $\tilde{A}$
in the $K$th row or those in the $K$th
column can decrease after node $k$ is removed. However, 
as opposed to the formulation of the Res strategy (\SEC\ref{sub:Restrepo}),
the removal of node $k$ does not result in
$\tilde{u}_K=0$, unless node $k$ is the only node contained in $M_K$.
Although this situation occurs after some nodes have been removed,
it is not very common except near the percolation threshold.
Therefore, we assume that the node removal changes the Perron vector to
\begin{equation}
\tilde{\bm u}+\Delta\tilde{\bm u}=\tilde{\bm u}+
\Delta^{\prime}\tilde{\bm u}-x\hat{e}_K,
\label{eq:Delta u coarse}
\end{equation}
where 
$\Delta^{\prime}\tilde{\bm u}$ is a small vector.
We determine $x$ as follows.
The $K$th linear equations for the Perron vector
before and after the removal of node $k$
are represented by
\begin{equation}
\sum_{I=1}^{N_{\rm M}} \tilde{A}_{KI}\tilde{u}_I = 
\tilde{\lambda}\tilde{u}_K
\end{equation}
and
\begin{equation}
\sum_{I=1,I\neq K}^{N_{\rm M}} \left(\tilde{A}_{KI}-d_{kI}\right)
\left(\tilde{u}_I+\Delta^{\prime}\tilde{u}_I-x\delta_{K,I}\right) = 
\left(\tilde{\lambda}+\Delta\tilde{\lambda}\right)
\left(\tilde{u}_K+\Delta^{\prime}\tilde{u}_K-x\right),
\end{equation}
respectively.
By combining these equations and neglecting
small-order terms 
$\Delta^{\prime}\tilde{u}_I\ll \tilde{u}_I$ ($1\le I\le N_{\rm M}$) and
$\Delta\tilde{\lambda}\ll \tilde{\lambda}$, 
we obtain
\begin{equation}
  x=\frac{1}{\tilde{\lambda}}\sum_{I=1,I\neq K}^{N_{\rm M}}d_{kI}\tilde{u}_I.
\label{eq:x}
\end{equation}
If node $k$ is the last node in $M_K$ that is removed at
a certain value of $p$, \EQ\eqref{eq:x} becomes
$x=\tilde{u}_K$. This relation is consistent with the fact that
$\tilde{u}_K$ vanishes after the removal of node $k$.

By substituting \EQS\eqref{eq:Delta A coarse},
\eqref{eq:Delta u coarse}, and \eqref{eq:x} and 
$\tilde{A}\tilde{\bm u}=\tilde{\lambda}\tilde{\bm u}$
in $(\tilde{A}+\Delta\tilde{A})
(\tilde{\bm u}+\Delta\tilde{\bm u})=
(\tilde{\lambda}+\Delta\tilde{\lambda})
(\tilde{\bm u}+\Delta\tilde{\bm u})$,
we obtain the following expression as the first-order approximation:
\begin{eqnarray}
\Delta\tilde{\lambda} &=&
- \frac{(2\tilde{u}_K - x+\Delta^{\prime}\tilde{u}_K)
\sum_{I\neq K} d_{kI} \tilde{u}_I+ 
\tilde{u}_K\sum_I d_{kI}\Delta^{\prime}\tilde{u}_I}
{\sum_I \tilde{u}_I^2-x\tilde{u}_K+\sum_I\tilde{u}_I\Delta^{\prime}\tilde{u}_I}\nonumber\\
&\approx& - \frac{(2\tilde{u}_K - x)\sum_{I\neq K} d_{kI} \tilde{u}_I}
{\sum_I \tilde{u}_I^2}.
\label{eq:Ik comm}
\end{eqnarray}
On the basis of 
\EQS\eqref{eq:x} and \eqref{eq:Ik comm}, we sequentially remove node $k$
that maximizes $(2\tilde{u}_K - x)\sum_{I\neq K} d_{kI} \tilde{u}_I$.
We label this immunization strategy as the Mod strategy.

When there are many nodes in module $K$,
\EQS\eqref{eq:Delta u coarse} and \eqref{eq:x} imply
$x=$ $\tilde{u}_K\sum_{I=1,I\neq K}^{N_{\rm M}}d_{kI}\tilde{u}_I$ $/$
$\sum_{I=1, I\neq K}^{N_{\rm M}}
\tilde{A}_{KI}\tilde{u}_I$ $\ll \tilde{u}_K$.
Therefore, the contribution of the node removal to $\Delta\tilde{\lambda}$
is attributed to two factors: the importance of the module that node $k$
belongs to (\textit{i.e.}, $2\tilde{u}_K-x\approx 2\tilde{u}_K$) and
the connectivity of node $k$ to
other important modules (\textit{i.e.}, 
$\sum_{I\neq K} d_{kI} \tilde{u}_I$).
As the other extreme to the case described above,
we consider the situation in which
node $k$ is the only node that
constitutes module $K$.
By substituting $d_{kI}=\tilde{A}_{KI}$ ($1\le I\le
N_{\rm M}$) in \EQS\eqref{eq:Delta u coarse}, \eqref{eq:x}, and
\eqref{eq:Ik comm}, we have
$\Delta\tilde{\lambda}=\tilde{u}_K^2/\sum_I \tilde{u}_I^2$;
the Res strategy is reproduced.
In other words, the Mod strategy is equivalent to the Res strategy when
all the nodes form isolated modules.

To apply the Mod strategy
to real data, we first partition the network into $N_{\rm M}$
modules. Then, we calculate $\tilde{u}_I$ ($1\le I\le N_{\rm M}$) by the
power method. 
This operation is fast unless
the spectral gap of $\tilde{A}$ is too small and
$N_{\rm M}$ is too large.
The power method produces $\tilde{\lambda}$ as a byproduct;
this value is used in
\EQ\eqref{eq:x}. Then, we remove the node that 
realizes the maximum
$(2\tilde{u}_K - x)\sum_{I\neq K} d_{kI} \tilde{u}_I$.
Next, we repeat this procedure.  To save
computation time, we do not apply a module detection algorithm
in each step. 
On the basis of the modular structure determined for the original network,
we recalculate $\tilde{u}_I$ and remove the nodes one at a time.
If all the modules are isolated,
we sequentially remove the nodes in the descending order of
$d_{kK}$. We recalculate $d_{kK}$ of all the remaining nodes
after the removal of each node.
This part of the Mod strategy
is heuristic and can be replaced by
other immunization strategies.

\section{Results}\label{sec:results}

We compare the efficiency of the Mod strategy on various networks with
those of other immunization strategies.

To detect modules in networks, we apply either the greedy algorithm
proposed by Clauset and colleagues
\cite{Newman04pre_fast,Clauset04pre} that approximately maximizes the
modularity of a network, the fast heuristic algorithm
to the same end proposed by 
Blondel and colleagues \cite{Blondel08jsm}, or the
algorithm based on random walks proposed by Rosvall and Bergstrom
\cite{Rosvall08pnas}. For all 
the examined data sets, Blondel's and Rosvall's algorithms
identify the smallest and the largest number of modules among the three
algorithms, respectively
(\TAB\ref{tab:statistics of networks}).  We call the Mod
strategy combined with the community detection algorithms of Clauset,
Blondel, and Rosvall as the Mod-C, Mod-B, and Mod-R strategies,
respectively.

We compare the efficiency of the Mod strategy
with the following immunization
strategies.

\begin{itemize}

\item \textit{Degree-based (D)
strategy}: We remove the nodes in decreasing order of their
degree in the original network. If there exists more than one node with
the same degree, we select one of them with equal probability.

\item \textit{Recalculated degree-based (RD)
strategy}: We sequentially remove the nodes with the largest degree.
This strategy differs from the D strategy in that we
recalculate the
degrees of all the remaining nodes after removing each node.

\item \textit{Betweenness-based (B) strategy}:
We remove the nodes in decreasing order of the betweenness
centrality. The betweenness centrality of a node is the normalized 
number of shortest paths between node pairs
that pass through the node \cite{Freeman79,Wasserman94,Boccaletti06}.

\item \textit{Recalculated betweenness-based (RB) strategy}:
We sequentially remove the nodes with the largest betweenness centrality.
We recalculate the betweenness centralities of all the remaining nodes
after removing each node.

\item \textit{Strategy based on dynamical importance (Res)}:
See \SEC\ref{sub:Restrepo} for the explanation \cite{Restrepo08prl}.

\end{itemize}

If, in any strategy, there are multiple nodes that realize
the maximum value of the relevant quantity, 
we select one of these nodes with equal probability.

Because the B strategy performs poorly compared to other strategies in all
the networks described in the following sections, we do not show the 
numerical results of this strategy.
Although the D strategy performs worse than the RD strategy (and many other
strategies) in most cases, we present the results obtained from the D
strategy because it is a typical strategy
\cite{Callaway00prl,Albert00nat,Cohen01prl}. While efficiencies of
the D, RD, B, and RB strategies are were compared in a previous study for
some networks \cite{Holme02pre_attack}, we examine these strategies
with regard to modular networks.

\subsection{Results for model networks}

\begin{figure}
\begin{center}
\includegraphics[width=6cm]{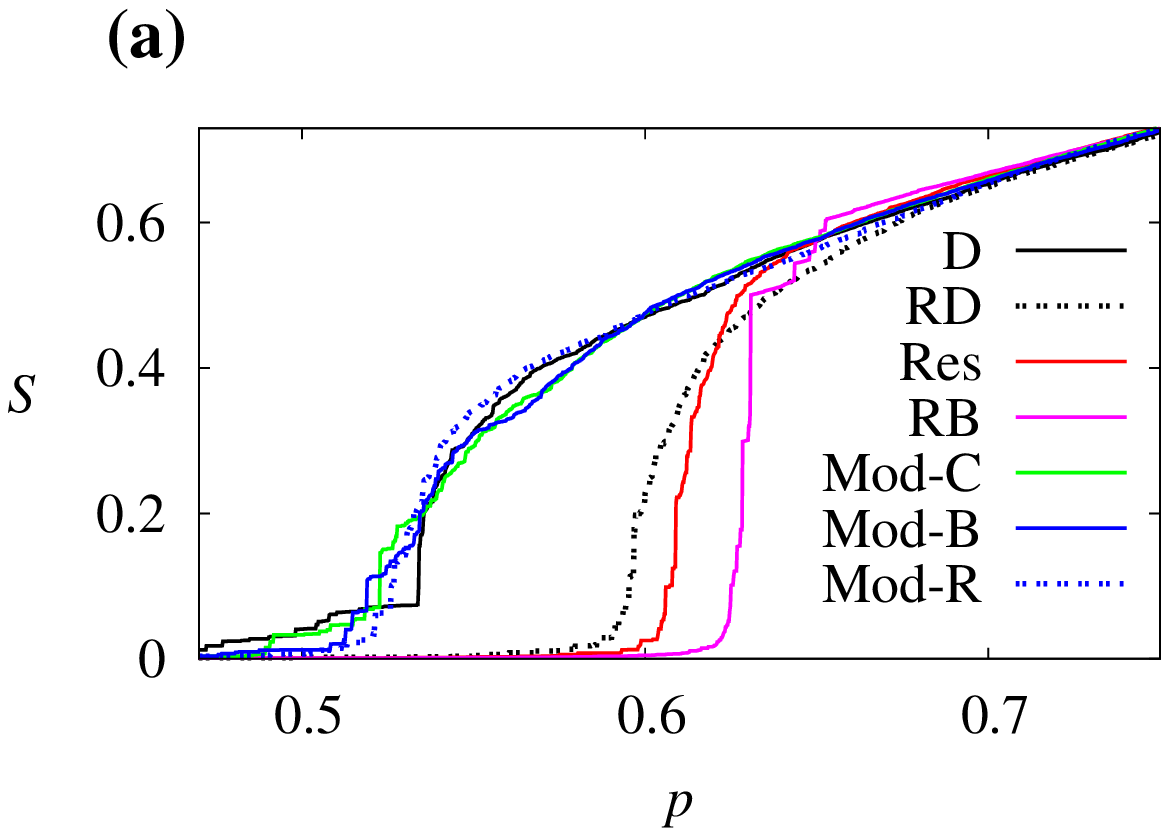}
\includegraphics[width=6cm]{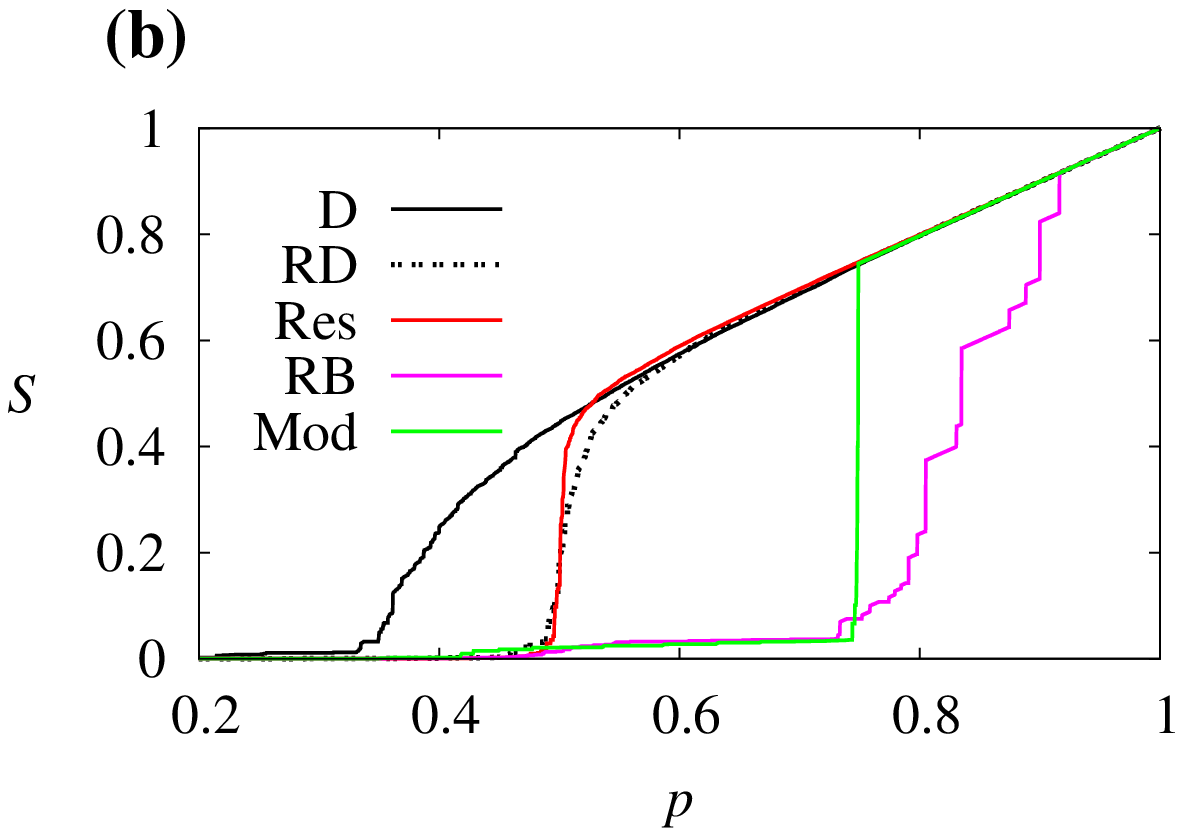}
\includegraphics[width=6cm]{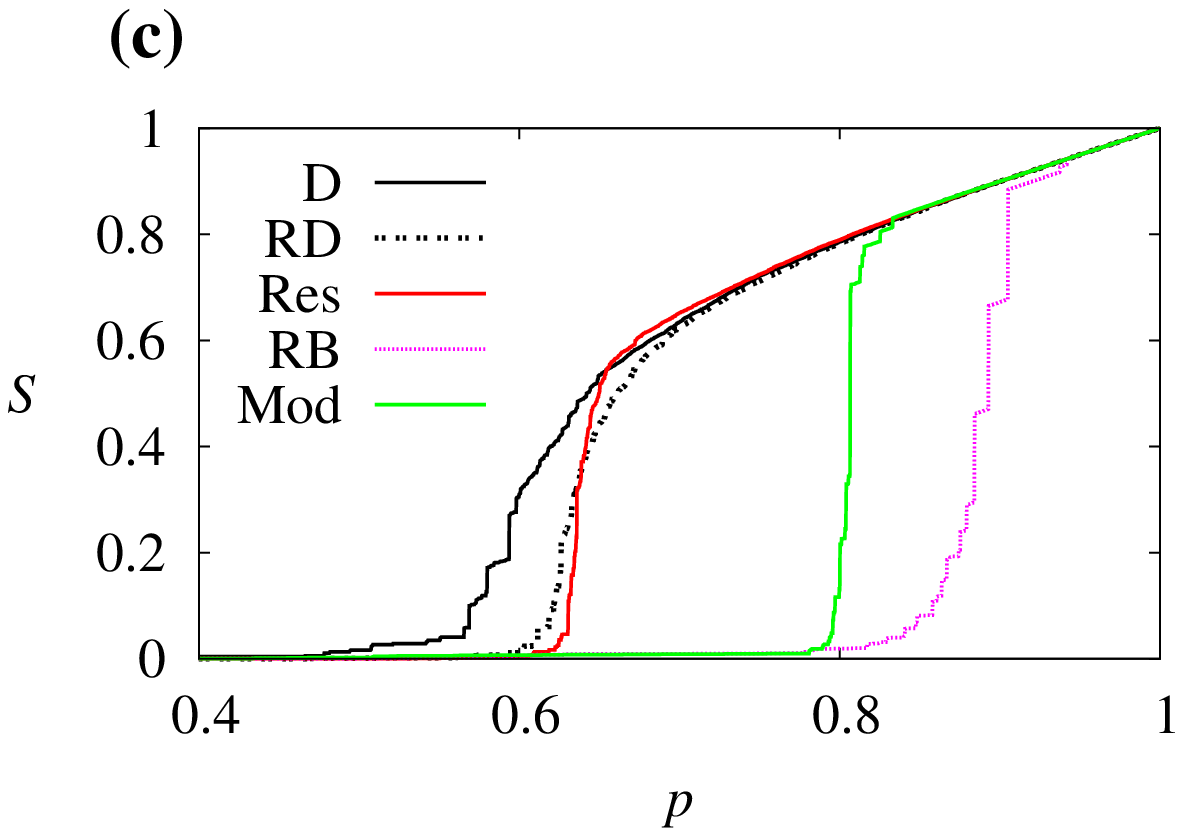}
\caption{Performance of different immunization strategies
in model networks with $N=5000$.
(a) Scale-free network.
(b) Ad hoc random network with $N_{\rm M}=25$ communities.
(c) Ad hoc scale-free network with $N_{\rm M}=100$ communities.}
\label{fig:adhoc}
\end{center}
\end{figure}

Our methods do not improve upon the previous methods for networks
without modular structure.  To verify this, we generate a scale-free
network with $N=5000$ and the degree distribution $p(k)\propto k^{-3}$
using the Barab\'{a}si-Albert (BA) model \cite{Barabasi99sci}.  We set
$\left<k\right>\approx 12$ by setting the parameters $m_0$ and $m$ of
the BA model to 6 \cite{Barabasi99sci}. Major statistics for the
generated BA model are listed in \TAB\ref{tab:statistics of
networks}. The relative size of the LCC is plotted against the node
occupation probability $p$ in \FIG\ref{fig:adhoc}(a). If $S$ is very
small for a large value of $p$, an immunization strategy is considered
to be efficient. The Mod-C, Mod-B, and Mod-R strategies are as
efficient as the D strategy.  These three strategies are superceded by
the RD, Res, and RB strategies, as expected.

The inefficiency of the Mod strategies is presumably caused by
the lack of the modular structure in the BA model.  In general, a
large Q-value indicates the presence of modular structure in a network
\cite{Newman04epjb,Newman04pre_fast,Fortunato09} (but see
\cite{Guimera04pre}).  The Q-values of this network
determined by the three community detection algorithms are equal to
0.249 (Clauset), 0.258 (Blondel), and 0.184 (Rosvall) and are
considered to be small. For a systematic comparison, we compare these
Q-values with those of the networks generated by random rewiring of
edges with the degree of each node preserved. The generated networks
do not have particular structure except that the degrees are
heterogeneous. The Q-values of the rewired networks are almost the
same as those of the BA model (\TAB\ref{tab:statistics of networks}),
which indicates the absence of modular structure in the BA model.

Next, we apply the Mod strategy to ad hoc networks with modular structure.
There are various algorithms that
produce benchmark networks with modular structure
\cite{Girvan02,Fortunato09}. We generate
two networks as follows. The following
numerical results do not critically depend on
the method of construction of the modular network.

Consider $N_{\rm M}$ modules of the same size $N/N_{\rm M}$.  In the first ad hoc
network, a module is the Erd\H{o}s-R\'{e}nyi 
random graph with the connection
probability $p_{\rm \ell}=\left<k\right>_{\rm \ell}/(N/N_{\rm M}-1)$, such
that the mean degree within a module is equal to $\left<k\right>_{\rm
\ell}=8$.  Then, we generate the coarse-grained network among $N_{\rm M}$
modules in the form of the random graph with a mean degree of
6. Any pair of node $i$
in module $M_I$ and node $j$ in another module $M_J$ ($J\neq I$)
may be connected if $M_I$ and $M_J$
are connected in the coarse-grained network. When this is
the case, we connect nodes $i$ and $j$ with probability
$\left<k\right>_{\rm g}/(6N/N_{\rm M})$. Then, 
for each node, the expected number of 
neighbors in different modules is equal to
$\left<k\right>_{\rm g}=1$.
We set $N=5000$ and $N_{\rm M}=25$ and run the algorithm
until we obtain a connected network.
The mean degree of the generated network
is equal to $\left<k\right>=8.82\approx
\left<k\right>_{\rm \ell}+\left<k\right>_{\rm g}$.

The results for different immunization strategies are
compared in \FIG\ref{fig:adhoc}(b).
The results labeled as Mod in \FIG\ref{fig:adhoc}(b)
are based on the predefined modular structure with
the number of modules $N_{\rm M}=25$, because
all the three algorithms for community detection
identify the correct modular
structure.
Figure~\ref{fig:adhoc}(b) indicates that
the Mod strategy substantially outperforms the Res strategy.
This is presumably because the Res strategy
detects LCCs contained in a single
module or a small number of modules as a signature
of a global epidemic, as discussed in \SEC\ref{sub:localized},
whereas the Mod strategy does not.

Figure~\ref{fig:adhoc}(b) indicates that the RB strategy outperforms
the Mod strategy.  This is as expected because a link version of the
RB strategy is used to partition the network efficiently into modules;
if we remove links in the decreasing order of the recalculated
betweenness centrality of the links, the network is partitioned into
modules efficiently \cite{Girvan02}. The drawback of the RB strategy
with respect to the Mod strategy is the former's high computation time;
we cannot apply the RB strategy to larger networks. We discuss this
point in \SEC\ref{sec:discussion}.

We also carry out numerical simulations on a heterogeneous ad hoc
modular network. We generate each module using the BA model with
$\left<k\right>_{\rm \ell}\approx 8$ (\textit{i.e.}, $m=m_0=4$).  The
coarse-grained network among modules is assumed to be the BA model
with a mean degree of 6 (\textit{i.e.}, $m=m_0=3$).  Pairs of nodes in
different modules are connected in the same way as in the previous
network, such that $\left<k\right>_{\rm g}=1$.  We set $N=5000$ and
$N_{\rm M}=100$.  The mean degree of the generated network
$\left<k\right>=8.59\approx \left<k\right>_{\rm
\ell}+\left<k\right>_{\rm g}$. The generated network is a connected
network.  The immunization results for this modular scale-free network
are shown in \FIG\ref{fig:adhoc}(c).  The results are qualitatively
the same as those in \FIG\ref{fig:adhoc}(b).

\subsection{Results for real-world networks}

We investigate the application of the Mod strategy to
four real-world networks. The statistics for each network
including the number of modules and the Q-values
are listed in \TAB\ref{tab:statistics of networks}.
The first example is
a high energy particle (HEP) citation
network \cite{hep}.
We use this network 
as a representative of a relatively dense network. This network is
used in a previous study of immunization \cite{Chen08prl}.
Because of its large mean degree, a relatively large fraction of nodes
have to be removed to fragment this network.
The Q-value for the partition using
the three algorithms are large. They are also much larger
than the Q-values for the networks generated by rewiring
the edges without changing the degree of each node.
Therefore, the HEP network has major modular
structure.

Note that the rewiring sometimes makes the network disconnected.
However, the Q-value does not differ much between
connected and disconnected rewired networks. Therefore, we do not
explore the effect of disconnectedness of the rewired networks.
We do the same omission for the three
other real-world networks examined later.

\begin{figure}
\begin{center}
\includegraphics[width=6cm]{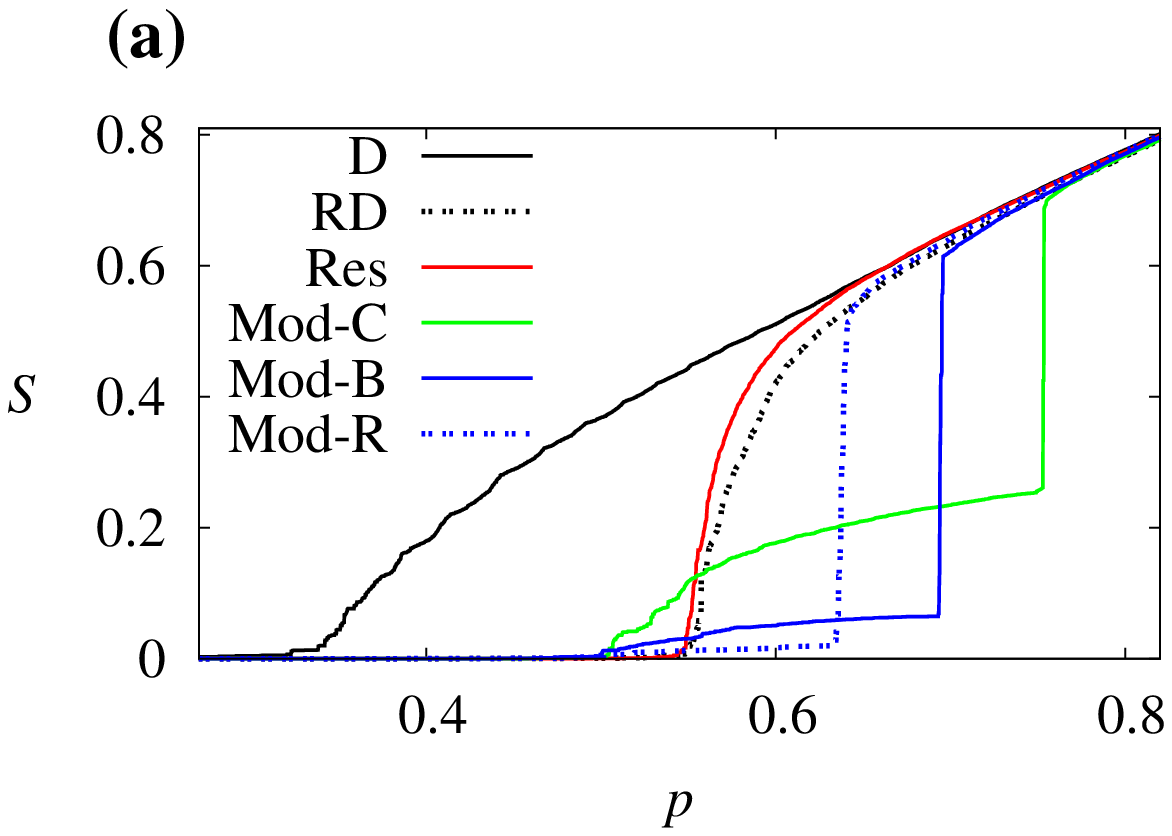}
\includegraphics[width=6cm]{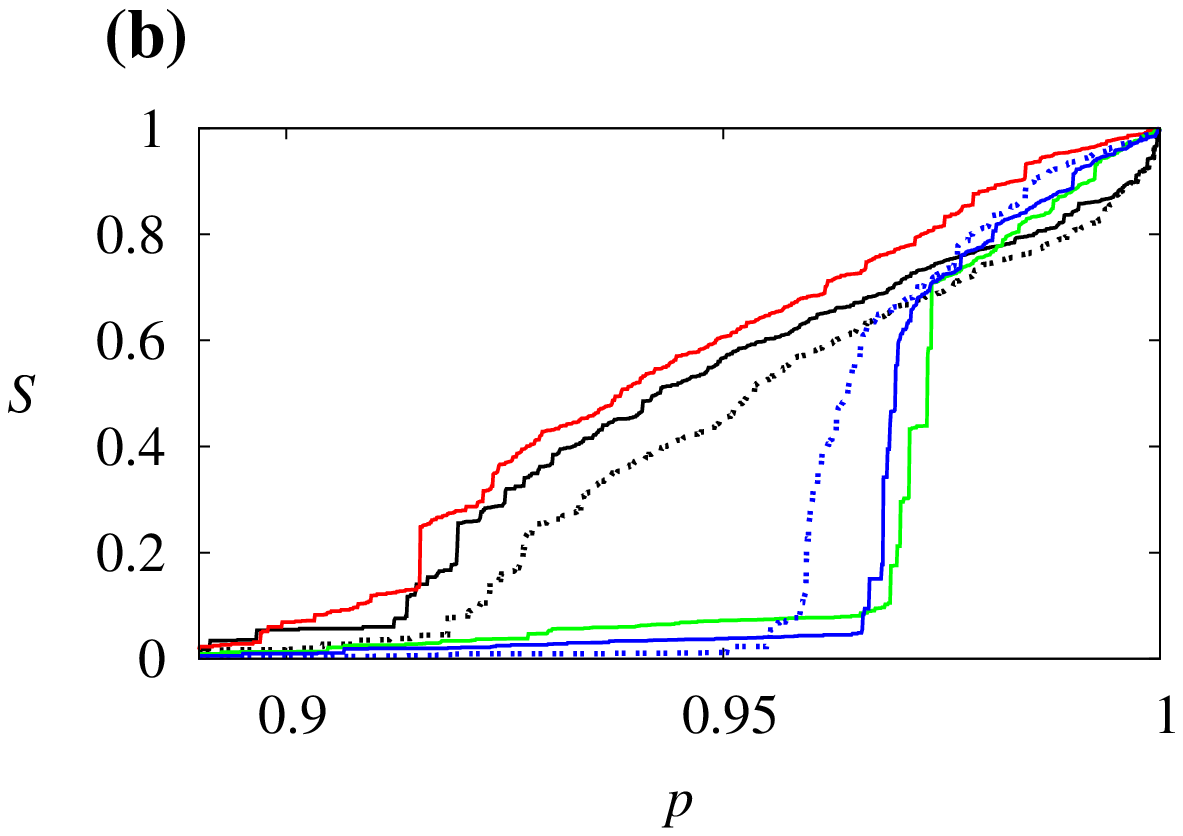}
\includegraphics[width=6cm]{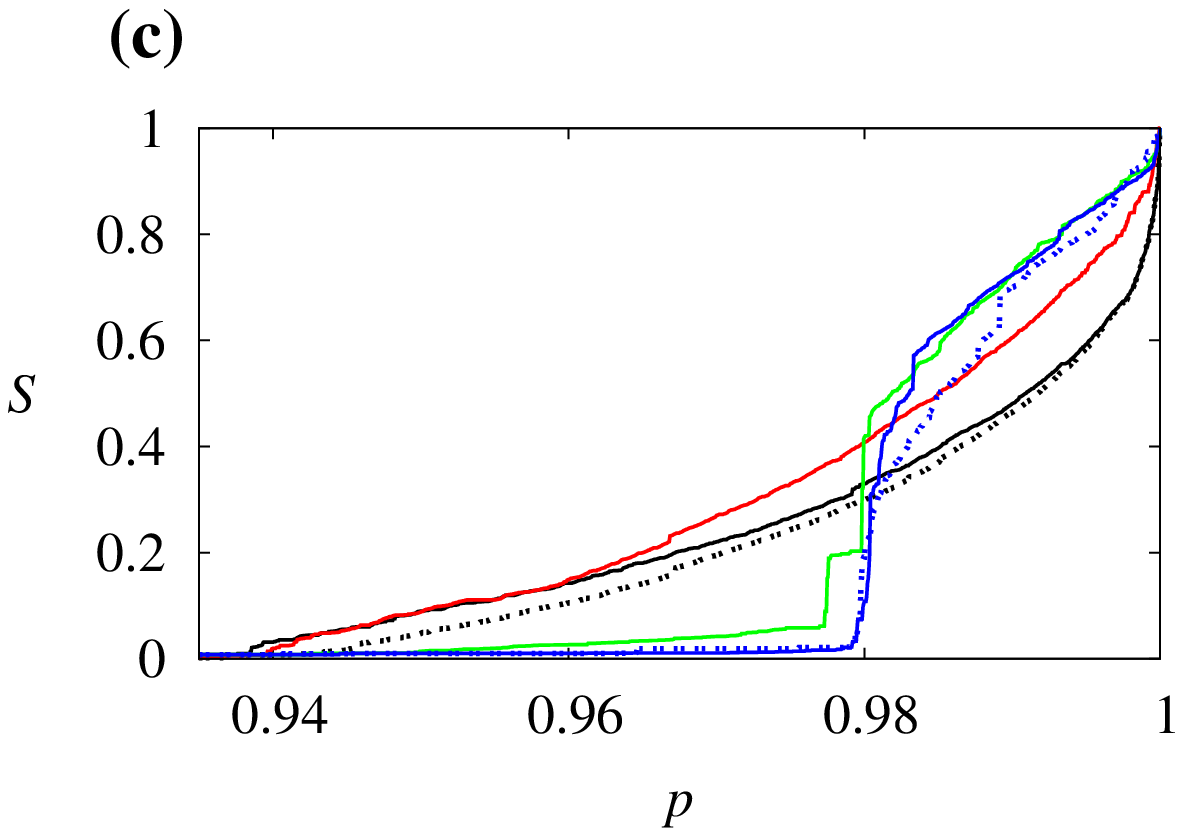}
\includegraphics[width=6cm]{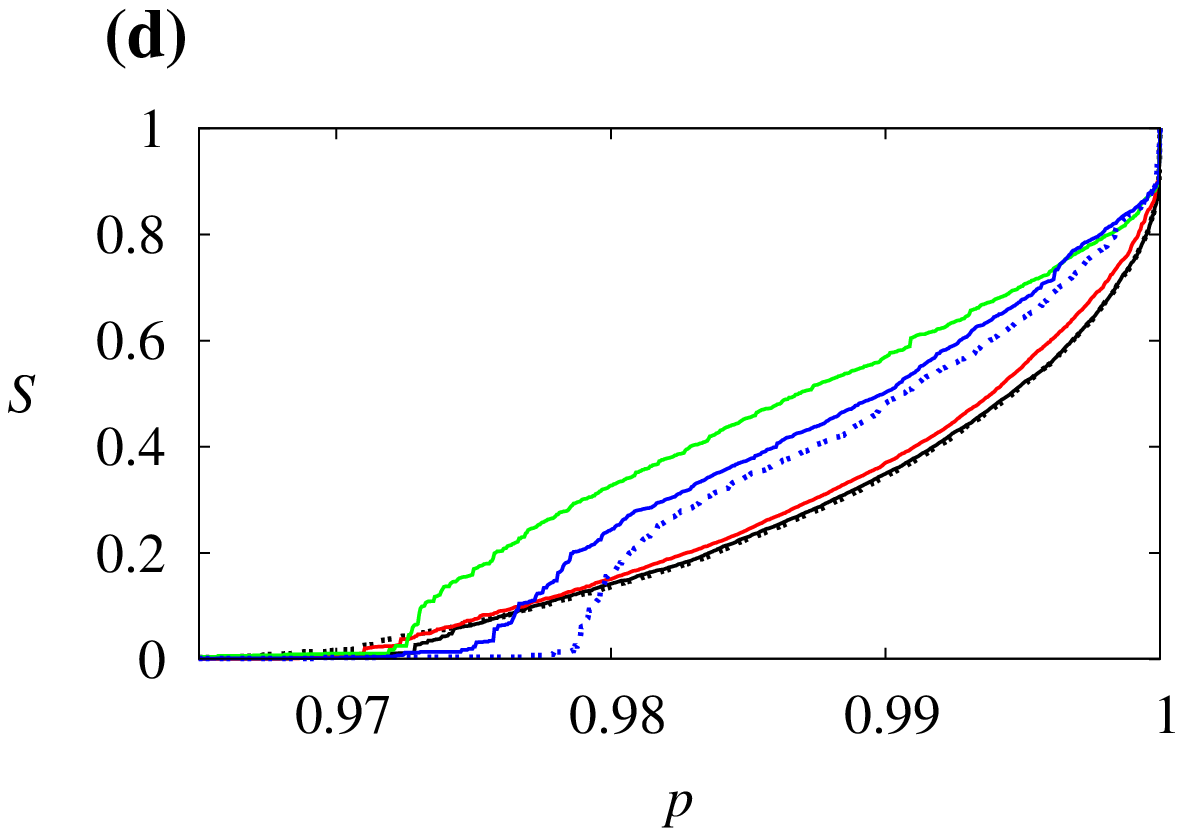}
\caption{Performance of different immunization strategies in
real networks.
(a) HEP collaboration network \cite{hep}.
(b) PGP social network \cite{Boguna04pre}.
(c) WWW \cite{Albert99}. (d) Email-based 
social network \cite{Ebel02_email}. In (d),
the results for the R strategy and those for the RD strategy 
overlap almost completely.}
\label{fig:real}
\end{center}
\end{figure}

The immunization results for the HEP network are shown in
\FIG\ref{fig:real}(a).
The results for the RB strategy are not shown because 
$N$ is too large for us to employ
the RB strategy. This limitation with regard to the RB strategy
is also true for the three other
networks. It can be observed from
\FIG\ref{fig:real}(a) that the Mod-R strategy
outperforms all the other
strategies including the Res strategy. The improvement obtained by 
employing the Mod-R strategy, which is
quantified
by the amount of shift of the percolation threshold
is approximately as large as
that obtained from the recently proposed strategy using
graph partitioning \cite{Chen08prl}.
This strategy \cite{Chen08prl}
divides the network into
equal-sized groups; it is distinct from the Mod strategy.

The LCC for the Mod-C strategy is small when
$p$ is large. However, below $p\approx 0.76$,
$S$ decreases slowly with a decrease in
$p$. At $p\approx 0.76$, all the modules are already separated. 
The LCC for the Mod-C strategy
occupies a significant fraction of the original network
at $p\approx 0.76$ and is represented by the largest module in the network.
Because we have not optimized the Mod strategy
after all the modules are separated,
the Mod-C strategy does not perform well below $p\approx 0.76$.
The Mod-B strategy yields a similar result;
below $p\approx 0.69$, the LCC is the largest module in the network.
However, the LCC
is smaller than that for the Mod-C strategy because the size 
of the largest module detected by Blondel's algorithm is smaller
than that detected by Clauset's algorithm.
The performance of the Mod strategies
can be enhanced if we
improve the immunization
strategy after all the modules are separated.
However, we do not
explore this aspect in the present study.

The second example is a social network called the 
Pretty Good Privacy (PGP) network \cite{Boguna04pre}. A
link is formed when
two persons share confidential information
using the PGP encryption algorithm on the Internet.
This network has a prominent community structure
(see \TAB\ref{tab:statistics of networks} for the Q-values).
The immunization results are shown in
\FIG\ref{fig:real}(b). For this network,
the Mod-C, Mod-B,
and Mod-R strategies outperform the D, RD, and Res strategies.

The third example is the LCC of a dataset of the World Wide Web 
\cite{Albert99}. We ignore the direction of the links.
The numerical results for this LCC are shown in
\FIG\ref{fig:real}(c). The Mod-C, Mod-B, and Mod-R strategies
perform better than
the D, RD, and Res strategies
at least in terms of the percolation threshold.

The fourth example is an email-based social network \cite{Ebel02_email}.
The results shown in \FIG\ref{fig:real}(d) indicate that, for this network,
the Mod-C, Mod-B, and Mod-R strategies do not 
outperform the other strategies.
The performance of the Mod-R strategy
is superior to those of the other methods near the percolation
threshold, but this superiority is only marginal.  
The performance of the Mod-C strategy is inferior to those of 
the other methods over the entire range of $p$.
The Mod-B and Mod-R strategies are more inefficient
than the D, RD, and Res strategies when $p$ is large.

The three community detection algorithms
result in large Q-values for the email social network. However,
this network may not be as modular as indicated by the 
large Q-values for two reasons. First, the rewired networks also have
relatively large Q-values,
although they are significantly smaller than the Q-values for the
original network (\TAB\ref{tab:statistics of networks}).
Second, generally speaking,
networks with small mean degree tend to have
large Q-values even if the modular structure is absent
\cite{Guimera04pre}.
The email social network may not have sufficient modular
structure, which may have caused the inefficiency of the Mod
strategy for this network.

\section{Discussion and Conclusions}\label{sec:discussion}

We have proposed an efficient algorithm called the Mod strategy
for immunizing
networks with modular structure. This strategy
combines a community detection
algorithm and the identification of nodes with crucial intermodular
links. We have validated the effectiveness of the Mod strategy
with artificial
and real networks using two community detection algorithms.  The Mod
strategy is applicable to networks in which the size of modules is
heterogeneous, as is the case in real modular networks \cite{Fortunato09}.

The Mod strategy can be extended to the case of networks with
more than two hierarchical levels, which
are often found in real data \cite{Ravasz02,Ravasz03}.
In such a network, we first remove 
the nodes responsible for the formation of
the most global connection. If modules
at the most global level have been fragmented, we apply the community
detection algorithm to each module 
such that the nodes responsible for connecting different
submodules in a module are preferentially removed.

For networks with bipartite modular structure
\cite{Fortunato09}, the Mod strategy
is inefficient.  This is because the Mod strategy is based on
the conventional concept of modular structure, \textit{i.e.}, there
are relatively more links within a module than across different
modules.  This property is not satisfied by networks with bipartite
modular structure. Dealing with bipartite modular structure
and also overlapping modular structure (see \cite{Fortunato09} for a
review) is beyond the scope of the present paper.

The Mod strategy does not outperform the RB strategy. This is as
expected because the RB strategy provides a useful algorithm for
community detection \cite{Girvan02}. The heart of the algorithm lies
in fragmenting a network into modules with a small number of links
(not nodes) that are removed in the decreasing order of the
betweenness centrality. However, carrying out community detection on
the basis of the RB strategy \cite{Girvan02} is computationally
formidable; this strategy requires ${\rm O}(N^3)$ time for sparse
networks. This fact has led to the development of faster algorithms
for community detection that are independent of the recalculated
betweenness centrality
\cite{Newman04epjb,Fortunato09,Newman04pre_fast,Clauset04pre,Blondel08jsm}.  The RB
strategy of immunization also requires ${\rm O}(N^3)$ time. An immunization
strategy developed by the adaptation of a faster community detection
algorithm that sequentially removes links (so-called divisive
algorithms) would outperform the Mod strategies
examined in the present study (\textit{i.e.}, Mod-C, Mod-B, and Mod-R).
However, such a community detection algorithm seems to be unknown
\cite{Fortunato09}.  We state that the Mod strategy outperforms the RB
strategy when the network is large. We have implicitly used fast
community detection algorithms so that the Mod strategy performs
faster than the RB strategy. For sparse networks, Rosvall's algorithm
runs comfortably fast.
Clauset's
algorithm runs faster than Rosvall's algorithm on our data and
requires only ${\rm O}(N \log^2 N)$ time \cite{Clauset04pre}.
Blondel's algorithm is even faster in general \cite{Blondel08jsm}.

In the so-called out-of-the-neighborhood (OUT) immunization
strategies \cite{Holme04epl}, one picks a neighbor of node that has
largest degrees out of the neighborhood of the original node. This is
an efficient immunization strategy that uses only the local
information about the network. The ring vaccination \cite{Cohen03prl}
stands on a similar spirit. In contrast to these strategies, the Mod
strategy has an important limitation that one needs global information
about the connectivity among modules. Nevertheless, the Mod and OUT
strategies are complementary with regard to the
information needed for implementation.  The Mod strategy
requires coarse but global information about the network, plus the
degree of each node. The OUT strategies require only the local
information about the network, but with the information about the
degree of the neighbors included.

Our results are consistent with the finding that nodes in a network
can be classified according to their global and local roles
\cite{Guimera05pnas,Guimera05nat,Guimera05jsm}.  This is particularly
true when the betweenness centrality is not predicted from the degree
\cite{Guimera05pnas}, which is typical for modular networks.  The
deviation of the global importance of a node from the local importance
of the same node in modular networks is also reported for the PageRank
and other similar centrality measures \cite{Masuda09njp,MKK09-1}. In
this situation, the Mod strategy preferentially immunizes globally
important nodes having important intermodular links rather than
locally important ones such as local hubs.  The general idea of
targeting globally important nodes in modular networks has potential
applications in other dynamical phenomena on networks, such as
epidemic dynamics, synchronization, opinion formation, and traffic.

\section*{Acknowledgments}

We thank Toshihiro Tanizawa and Taro Ueno for their valuable discussions.
N.M. acknowledges the support through
Grants-in-Aid for Scientific Research
(Nos. 20760258 and 20540382) from MEXT, Japan.

\newpage
\clearpage

\begin{figure}
\begin{center}
\includegraphics[width=6cm]{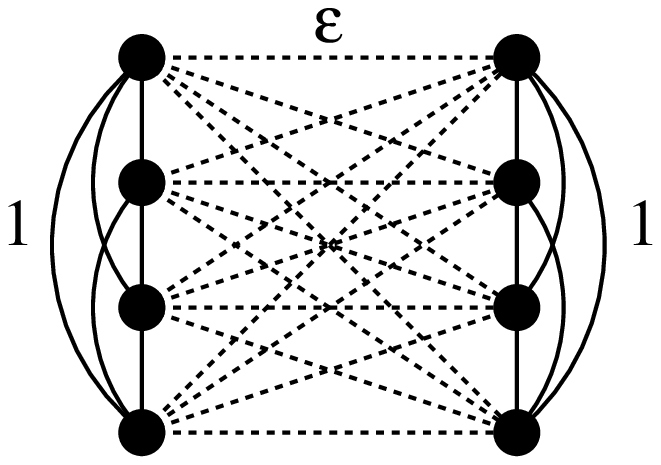}
\caption{Approximated modular network with
$N=8$ and $N_{\rm M}=2$. The solid and dotted lines
represent links with weights 1 and $\epsilon$, respectively.}
\label{fig:community anneal}
\end{center}
\end{figure}

\newpage
\clearpage

\newpage
\clearpage

\newpage
\clearpage

\begin{table}
\begin{center}
\caption{Statistics of networks. The number of nodes and links are those of
the LCC of the network. $N_{\rm M}$ is the number of 
modules detected by each algorithm. 
For the rewired networks, the average and the standard deviation of
the Q-values are shown for each community detection algorithm.
To this end, we generate 100 rewired networks from each original network.}
\begin{tabular}{|p{1.6cm}|p{2.0cm}|p{1.5cm}|p{1.5cm}|p{1.5cm}|p{1.5cm}|p{1.5cm}|}
\hline
\multicolumn{2}{|c|}{network} & BA & HEP & PGP & WWW & email\\ \hline
\multicolumn{2}{|c|}{number of nodes ($N$)} & 5000 & 27400 & 10680 & 99193 & 63495\\ \hline
\multicolumn{2}{|c|}{number of links} & 29979 & 352021 & 24340 & 198355 & 107689\\ \hline
& Clauset & 13 & 143 & 196 & 1079 & 310\\ \cline{2-7}
$N_{\rm M}$& Blondel & 12 & 29 & 99 & 210 & 121\\ \cline{2-7}
& Rosvall & 266 & 681 & 921 & 3511 & 2536\\ \hline
Q-value& original & 0.249 & 0.519 & 0.852 & 0.853 & 0.731\\ \cline{2-7}
(Clauset)& rewired & 0.250 & 0.148 & 0.466 & 0.480 & 0.520\\
& & $\pm$ 0.002 & $\pm$ 0.000 & $\pm$ 0.001 & $\pm$ 0.002 & $\pm$ 0.003\\ \hline
Q-value& original & 0.258 & 0.648 & 0.883 & 0.895 & 0.786\\ \cline{2-7}
(Blondel)& rewired & 0.258 & 0.156 & 0.470 & 0.499 & 0.562\\
& & $\pm$ 0.003 & $\pm$ 0.001 & $\pm$ 0.001 & $\pm$ 0.000 & $\pm$ 0.001\\ \hline
Q-value& original & 0.184 & 0.585 & 0.812 & 0.832 & 0.724\\ \cline{2-7}
(Rosvall)& rewired & 0.164 & 0.008 & 0.400 & 0.436 & 0.511\\
& & $\pm$ 0.007 & $\pm$ 0.001 & $\pm$ 0.001 & $\pm$ 0.000 & $\pm$ 0.000\\ \hline
\end{tabular}
\label{tab:statistics of networks}
\end{center}
\end{table}

\end{document}